\documentclass{article}

\usepackage{amssymb,cite} 

\def \BEA { \begin{eqnarray}}
\def \EEA {\end{eqnarray}}
\def \BE {\begin{equation}}
\def \EE {\end{equation}}
\def\d{\mathrm{d}}

\def \WDS #1 {\mbox{$\Phi_{#1}^{S}$}}
\def \WDA #1 {\mbox{$\Phi_{#1}^{A}$}}
\def \WD #1 {\mbox{$\Phi_{#1}$}}

\def \mi {\stackrel{i}{m}}
\def \mj {\stackrel{j}{m}}
\def \mk {\stackrel{k}{m}}
\def \mr {\stackrel{r}{m}}
\def \ms {\stackrel{s}{m}}
\def \mp {\stackrel{p}{m}}
\def \mz {\stackrel{z}{m}}
\def \mq {\stackrel{q}{m}}
\def \mo {\stackrel{o}{m}}
\def \mD {\stackrel{2}{m}}
\def \mT {\stackrel{3}{m}}
\def \mC {\stackrel{4}{m}}

\def \mio #1 {\mi_{#1}\ ^{  \! \! \! \! 0}} 
\def \mjo #1 {\mj_{#1}\ ^{  \! \! \! \! 0}} 
\def \mko #1 {\mk_{#1}\ ^{  \! \! \! \! 0}} 
\def \mro #1 {\mr_{#1}\ ^{  \! \! \! \! 0}} 
\def \mso #1 {\ms_{#1}\ ^{  \! \! \! \! 0}} 
\def \mpo #1 {\mp_{#1}\ ^{  \! \! \! \! 0}} 
\def \mzo #1 {\mz_{#1}\ ^{  \! \! \! \! 0}} 
\def \mqo #1 {\mq_{#1}\ ^{  \! \! \! \! 0}} 
\def \moo #1 {\mo_{#1}\ ^{  \! \! \! \! 0}} 
\def \mDo #1 {\mD_{#1}\ ^{  \! \! \! \! 0}} 
\def \mTo #1 {\mT_{#1}\ ^{  \! \! \! \! 0}} 
\def \mCo #1 {\mC_{#1}\ ^{  \! \! \! \! 0}} 

\def \miJ #1 {\mi_{#1}\ ^{  \! \! \! \! (1)}} 
\def \mjJ #1 {\mj_{#1}\ ^{  \! \! \! \! (1)}} 
\def \mkJ #1 {\mk_{#1}\ ^{  \! \! \! \! (1)}} 
\def \mrJ #1 {\mr_{#1}\ ^{  \! \! \! \! (1)}}

\def \bl {\mbox{\boldmath{$\ell$}}}

\def \bn {\mbox{\boldmath{$n$}}}

\def \hbm #1 {\mbox{\boldmath{$\hat m^{(#1)}$}}}
\def \bm {\mbox{\boldmath{$m$}}}

\def \Mk {\stackrel{k}{M}}

\newcommand{\be}{\begin{equation}}
\newcommand{\ee}{\end{equation}}
\newcommand{\beqn}{\begin{eqnarray}}
\newcommand{\eeqn}{\end{eqnarray}}
\newcommand{\pa}{\partial}
\newcommand{\ba}{\begin{array}}
\newcommand{\ea}{\end{array}}
\newcommand{\pp}{{\it pp\,}-}

\def \BEAH {\begin{eqnarray*}}
\def \EEAH {\end{eqnarray*}}
\def \BEA {\begin{eqnarray}}
\def \EEA {\end{eqnarray}}
\def \BDM {\begin{displaymath}}
\def \EDM {\end{displaymath}}

\def \Mk {\stackrel{k}{M}}

\def \pul {{{\footnotesize{\frac{1}{2}}}}}

\newcommand{\M}[3] {{\stackrel{#1}{M}}_{{#2}{#3}}}
\newcommand{\m}[3] {\!{\stackrel{\hspace{.3cm}#1}{m}}_{\!\!{#2}{#3}}\,}

\begin{document}

\title{Type III and N Einstein spacetimes in higher dimensions: \\ general properties}

\author{Marcello Ortaggio\thanks{ortaggio(at)math(dot)cas(dot)cz}, Vojt\v ech Pravda\thanks{pravda@math.cas.cz}, Alena Pravdov\' a\thanks{pravdova@math.cas.cz} \\
Institute of Mathematics, Academy of Sciences of the Czech Republic \\ \v Zitn\' a 25, 115 67 Prague 1, Czech Republic}
\date{\today}

\maketitle

\abstract{The Sachs equations governing the evolution of the optical matrix of geodetic WANDs (Weyl aligned null directions) are explicitly solved in $n$-dimensions in several cases which are of interest in potential applications. This is then used to study Einstein spacetimes of type III and N in the higher dimensional Newman-Penrose formalism, considering both Kundt and expanding (possibly twisting) solutions. In particular, the general dependence of the metric and of the Weyl tensor on an affine parameter~$r$ is obtained in a closed form. This allows us to characterize the peeling behaviour of the Weyl ``physical'' components for large values of $r$, and thus to discuss, e.g., how the presence of twist affects polarization modes, and qualitative differences between four and higher dimensions. Further, the $r$-dependence of certain non-zero scalar curvature invariants of expanding spacetimes is used to demonstrate that curvature singularities may generically be present. As an illustration, several explicit type N/III spacetimes that solve Einstein's vacuum equations (with a possible cosmological constant) in higher dimensions are finally presented.}

\section{Introduction}

It was recognized  long ago that several important features of gravitational radiation in General Relativity can be conveniently described in a covariant manner by studying asymptotic properties of spacetimes \cite{Sachs61,BBM,NP,Penrose63}. More generally, the development of asymptotic techniques has proven fundamental in understanding general properties of the theory, since the behaviour of the gravitational field near (spacelike or null) infinity encodes essential information about physical quantities such as mass, angular momentum and flux of radiation (at least in asymptotically flat spacetimes). From a technical viewpoint, the Newman-Penrose formalism \cite{NP} turns out to be extremely useful in analyzing the fall-off properties of gravitational fields at infinity. In a nutshell, {(after making certain initial technical assumptions)} from the Ricci identities one first extracts the $r$-dependence of some of the Ricci rotation coefficients which are needed in the analysis (throughout the paper $r$ will denote an affine parameter along null geodesics). Subsequently, specific Bianchi identities are integrated that determine the behaviour of the Riemann (Weyl) tensor, which geometrically characterizes properties of the gravitational field. From this one demonstrates, e.g., the characteristic peeling-off properties of radiative spacetimes \cite{Sachs61,BBM,NP,Penrose63}. Then one can further proceed to integrate the remaining Ricci/Bianchi identities and thus find asymptotic solutions (possibly with specific extra assumptions, see, e.g., \cite{NewUnt62}, and \cite{NewTod80} for a review and further references).

It was noticed in \cite{Sachs61} that four dimensional {\em algebraically special} spacetimes, while leading to significant mathematical simplification, still asymptotically retain the essential features of (outgoing) radiation fields generated by more realistic sources. In particular, in that case the $r$-dependence of the Weyl tensor can be determined in closed form (and not only asymptotically). This is also an important first step towards the exact integration of the full Newman-Penrose equations, aimed at determining the explicit metric functions. One can indeed get quite far in the case of algebraically special spacetimes, at least in vacuum \cite{RobTra62,Kerr63,Kinnersley69,RobRobZun69,DebKerSch69,Talbot69,Lind74}, and several important exact solutions fall within this large class (see \cite{NewTod80,Stephanibook} for further references).

In recent years the interest in gravity in higher dimensions has grown considerably, mainly motivated by modern unified theories, AdS/CFT and recent brane world scenarios. Notions such as the total energy of an isolated system and energy flux are thus fundamental also in higher dimensional theories \cite{HolIsh05,HolWal04,Ishibashi08}. The study of radiation in higher dimensions may ultimately enable one to distinguish different models, since properties of gravitational waves depend on the model under consideration (in particular, on spacetime dimensions \cite{CarDiaLem03}). It is thus now of interest to explore ideas and techniques similar to those mentioned above for the study of asymptotic properties of algebraically special spacetimes with $n>4$ dimensions. 
The necessary formalism has been provided in recent works, where an $n>4$ generalization of the Petrov classification \cite{Coleyetal04} and of the Newman-Penrose equations \cite{Pravdaetal04,Coleyetal04vsi,OrtPraPra07} have been presented.\footnote{Very recently, an extension of the GHP formalism to higher dimensions has also been developed \cite{Durkeeetal10}. Although we will not need the GHP formalism here, the results of \cite{Durkeeetal10} are useful also in the NP context, since some redundancy of the original Bianchi equations of \cite{Pravdaetal04} has been removed. Beware of the fact that some normalizations used in \cite{Durkeeetal10} slightly differ from those of the present paper.} Such a programme was thus started in \cite{OrtPraPra09b}, where we studied vacuum spacetimes admitting a non-degenerate (as defined in section~\ref{sec_Sachs}) geodetic multiple WAND, and satisfying a further condition necessary for asymptotic flatness. Thanks to the results of \cite{Pravdaetal04}, the assumed non-degeneracy implies that the only possible algebraic type of that family is II (or D), but not III and N. We observed that such spacetimes do not peel-off and do not contain gravitational radiation, as opposed to the case of four dimensions \cite{Sachs61}. 

The analysis of \cite{OrtPraPra09b} can now be extended in various directions by modifying some of the assumptions made there. It is the purpose of the present paper to focus on empty spacetimes of type III and N, with a possible cosmological constant.  For these, the (unique) multiple WAND is necessarily geodetic and degenerate \cite{Pravdaetal04}. We can already remark at this point that this implies that such spacetimes can not be asymptotically flat (even in a ``weak'', local sense, in contrast to the $n=4$ case).\footnote{Cf.~footnote 1 of \cite{OrtPraPra09b}.} Our analysis can be however still of interest for spacetimes with, e.g., Kaluza-Klein asymptotics. It is also worth observing that, as opposed to \cite{OrtPraPra09b}, we will not need here any extra assumptions on the asymptotics of the Weyl tensor -- its full $r$-dependence will be fixed by the Bianchi identities.\footnote{The extra assumption needed in \cite{OrtPraPra09b} concerned the asymptotic behaviour of the Weyl components $C_{ijkl}$ (in the notation of \cite{Coleyetal04}, see also the following), which vanish identically in the case of type III/N spacetimes.}

The paper is organized as follows. In section~\ref{sec_Sachs} the Sachs equations for a congruence of geodetic WANDs $\bl$ are studied for spacetimes that satisfy the (rather weak) condition $R_{00}\equiv R_{ab}\ell^a\ell^b=0$, and an explicit solution is given when the principal directions of shear and twist are aligned. This includes several cases of interest and, in particular, Einstein spacetimes of type N and III, which are then studied in the rest of the paper. Namely, in sections~\ref{sec_N} and \ref{sec_III} we  determine the $r$-dependence of the Weyl tensor components of such spacetimes in a parallelly transported frame. We can thus discuss, e.g., their peeling behaviour near infinity, curvature invariants and possible singularities, frame freedom and rotation of frames induced by the presence of twist. Differences with respect to the four-dimensional case are also pointed out. In the final appendix we construct several examples of Einstein spacetimes of type N and III. These are explicitly given in five dimensions, but they can also be easily extended to higher dimensions if desired.

\paragraph{Notation} 

Following \cite{Coleyetal04,Pravdaetal04,Coleyetal04vsi,OrtPraPra07}, we use a frame consisting of two null vectors $\bm_{(0)}=\bl$  (aligned with the multiple WAND) and $\bm_{(1)}=\bn$, and $n-2$ orthonormal spacelike vectors $\bm_{(i)}$,  
where $i, j, \dots=2,\ldots,n-1$. In terms of these, the metric reads
\be
 g_{ab}=2l_{(a}n_{b)}+\delta_{ij}m^{(i)}_am^{(j)}_b , \label{metricIII}
\ee 
where, hereafter, $a,\ b=0,\ 1,\dots,\ n-1$.

Derivatives along the frame vectors $\bl$, $\bn$ and $\bm_{(i)}$ are denoted by $D$, $\Delta$ and $\delta_i$, respectively. We choose the frame such that it is {parallelly transported} along $\bl$. The optical matrix $L$ of $\bl$ has matrix elements
\be
 L_{ij}=\ell_{a;b}m_{(i)}^am_{(j)}^b , 
\ee  
with (anti-)symmetric parts 
\be
 S_{ij}=L_{(ij)} , \qquad A_{ij}=L_{[ij]} . 
\ee
The optical scalars expansion, $\theta $, shear, $\sigma$, and twist, $\omega$, are defined by $\theta =L_{ii}/(n-2)$,
$\sigma^2=(S_{ij}-\theta \delta_{ij})(S_{ij}-\theta \delta_{ij})$, and $\omega^2=A_{ij}A_{ij}$. Other Ricci rotation coefficients used in this paper are defined by (see \cite{Pravdaetal04,OrtPraPra07} for the full set of coefficients)
\be
 L_{1i}=\ell_{a;b}n^a m_{(i)}^b , \qquad L_{i1}=\ell_{a;b} m_{(i)}^a n^b , \qquad L_{11}=\ell_{a;b}n^a n^b , \qquad \M{i}{j}{k}=-\M{j}{i}{k}=m_{(i)a;b}m_{(j)}^am_{(k)}^b .
\ee
The Weyl tensor of spacetimes of type III and N has only negative boost weight frame components, for which we use the compact symbols 
\be
  \Psi_{i} = C_{101i}, \qquad  \Psi_{ijk}= \frac{1}{2} C_{1kij}, \qquad \Psi_{ij} = \frac{1}{2} C_{1i1j} .
\ee

From the symmetries and the tracelessness of the Weyl tensor one has the identities \cite{Pravdaetal04} 
\be 
 \Psi_i=2 \Psi_{ijj}, \qquad  \Psi_{\{ijk\}}=0, 
 \qquad  \Psi_{ijk}=-\Psi_{jik}, \qquad  \Psi_{ij}=\Psi_{ji}, \qquad  \Psi_{ii}=0,
\ee 
{where $\Psi_{\{ijk\}}\equiv \Psi_{ijk}+\Psi_{kij}+\Psi_{jki}$}, which will be employed throughout the paper. A subscript or superscript $0$ will denote quantities which are independent of $r$ (e.g., $s_{(2)}^0$, $a_0$, $\Psi^0_{ijk}$, etc.).

\section{Geodetic WAND: solving the Sachs equations}

\label{sec_Sachs}

We are interested in studying asymptotic properties of spacetimes along the congruence generated by a geodetic multiple WAND $\bl$. As a first step, it is thus natural to fix the $r$-dependence of the matrix $L_{ij}$, which determines the optical properties of $\bl$. This can be done by integrating the Sachs equations (a subset of the Ricci identities), which for a geodetic, affinely parametrized WAND (not necessarily multiple) read \cite{OrtPraPra07}
\be
 DL_{ij}=-L_{ik}L_{kj} ,
 \label{sachs}
\ee
where we assumed that the Ricci tensor satisfies $R_{00}=0$ (which obviously holds for Einstein spaces, defined by $R_{ab}=g_{ab}R/n$). When $L_{ij}=0$ we have a trivial solution of~(\ref{sachs}) corresponding to Kundt spacetimes. In the rest of this section we will study only the non-trivial case $L_{ij}\neq0$.

When $\bl$ is non-degenerate, i.e. $L$ is invertible, such a matrix differential equation can be easily solved in terms of $L^{-1}$ \cite{OrtPraPra09b}, and by taking the inverse matrix one then finds $L$. When the number of dimensions is kept arbitrary, {this is done more conveniently by} expanding $L$ in a power series in $1/r$. This was indeed the starting point in the analysis of \cite{OrtPraPra09b}. By contrast, here we will not assume $L$ to be invertible, but we will solve (\ref{sachs}) in a closed form under some other assumptions. These will be such to include the form of $L$ compatible, in particular, with type III/N spacetimes \cite{Pravdaetal04}. However, in this section we will be general enough so that the presented results will apply in a wider context.

\subsection{Explicit solution when the principal directions of shear and twist are aligned}

If one forgets for a moment about the request that the frame be parallelly propagated, one can always choose the basis vectors $\bm_{(i)}$ such that the symmetric or the antisymmetric parts of the matrix $L$ take their canonical form (that is, $S_{ij}$ is diagonal, or $A_{ij}$ is block-diagonal with 2-dimensional anti-symmetric blocks). It is natural to refer to such preferred basis vectors $\bm_{(i)}$ as the {\em principal shear directions} and {\em principal twist directions}, respectively (there may be some degeneracy, in general, i.e. there  need not be a unique basis of principal directions). 
Here we consider the special case when there exists a basis of vectors that are principal directions of shear and twist simultaneously, {which is relevant to important applications (and includes, in particular, the case when $L$ is a {\em normal matrix, i.e., $[S,A]=0$)}. This means that $L$ admits a canonical form given by a direct sum of 2-blocks of the form
\be
 {\cal L}=\left(\begin {array}{rr} s_{(2)} & A_{23} \\
 -A_{23} & s_{(3)}
\end {array}
 \right) ,
 \label{L_block}
\ee
where, for definiteness, the frame indices refer to the first block (next blocks will be characterized by pairs of indices $(4,5)$, $(6,7)$, \ldots). 
If the spacetime dimension $n$ is odd, there will be also an extra one-dimensional block.

\subsubsection{The canonical frame can be parallelly transported} 

\label{subsubsec_parall}

We want to solve eqs.~(\ref{sachs}), which hold when the basis vectors $\bm_{(i)}$ are parallelly transported along $\bl$. Since we are now assuming that the $\bm_{(i)}$ coincide with the common principal shear and twist directions, for consistency it is necessary to prove that such vectors can indeed be parallelly transported.

Let us thus take such a canonical frame at a special value of $r$, say $r=0$. Then we have, by construction, that  $L|_{r=0}$ is block-diagonal, with blocks of the form
\be
 {\cal L}|_{r=0}=\left(\begin {array}{rr} s^0_{(2)} & A^0_{23} \\
 -A^0_{23} & s^0_{(3)}
\end {array}
 \right).
 \label{L0_block}
\ee
One can now define a frame in a neighborhood of $r=0$ by parallel transporting the frame defined at $r=0$. 
By Taylor-expanding $L=L|_{r=0}+r(DL)|_{r=0}+\frac{1}{2}r^2(D^2L)|_{r=0}+\ldots$, the evolution of $L$ ({assumed to be analytic}) will be determined by (\ref{sachs}) and its $r$-derivatives 
\be
D^mL=(-1)^m m! L^{m+1} ,
\ee
{evaluated} at $r=0$.
Since $L^m|_{r=0}$ has clearly the same block-diagonal structure (for any $m$), it follows  that $(D^mL)|_{r=0}$ has the same block-form as well.
Consequently $L$ has such a block-diagonal form for any $r$, with blocks given by (\ref{L_block}), {that is what we wanted to prove}.

\subsubsection{Explicit $r$-dependence} 

Since now each 2-dimensional block obeys a decoupled equation, we have effectively reduced eq.~(\ref{sachs}) to the standard Sachs equations of $n=4$ gravity. For {the first} block, we thus have the general solution \cite{Sachs61,NP,penrosebook2} 
\beqn
 & &  
s_{(p)}=\frac{s_{(p)}^0+r\big[s_{(2)}^0s_{(3)}^0+(A_{23}^0)^2\big]}{1+r\big(s_{(2)}^0+s_{(3)}^0\big)+r^2\big[s_{(2)}^0s_{(3)}^0+(A_{23}^0)^2\big]}  \qquad (p=2,3), \label{s_aligned} \\
& & A_{23}=\frac{A_{23}^0}{1+r\big(s_{(2)}^0+s_{(3)}^0\big)+r^2\big[s_{(2)}^0s_{(3)}^0+(A_{23}^0)^2\big]} ,
\label{A_aligned}
\eeqn
and similarly for the next blocks. In odd dimensions, the last block will be 1-dimensional and given by $s_{(n-1)}=s_{(n-1)}^0/(1+rs_{(n-1)}^0)$.

The optical scalars are 
\be
 \theta=\textstyle{\frac{1}{n-2}}\sum_{i=2}^{n-1}s_{(i)} , \qquad \sigma^2=\sum_{i=2}^{n-1}(s_{(i)}-\theta)^2 , \qquad \omega^2=2(A_{23}^2+A_{45}^2+\ldots).
\ee

More special subcases are discussed below.

\subsubsection{Non-twisting case} 

The case $\omega=0$ is of course included in the above ``aligned'' case since $A_{ij}=0$. Then, $L_{ij}=S_{ij}$ is diagonal with eigenvalues (using (\ref{s_aligned}) with $A_{ij}^0=0$)
\be
 s_{(i)}=\frac{s_{(i)}^0}{1+rs_{(i)}^0} ,
\ee
as obtained in \cite{PraPra08}. 

Note that in this case $\bl$ must be expanding ($\theta\neq 0$) \cite{OrtPraPra07} or one is simply left with the Kundt class (see, e.g., \cite{Coleyetal03,Coleyetal06,PodZof09} {and references therein}).

\subsubsection{Non-shearing case} 

When $\sigma=0$ one has $S_{ij}=\theta\delta_{ij}$, which is clearly aligned with any $A_{ij}$. Then $s_{(i)}=\theta$ for any $i=2,\ldots,n-2$, so that by (\ref{s_aligned}) all non-zero $A_{ij}$ take the same value (up to a sign). All the information about $L_{ij}$ is thus contained in the two optical scalars
\be
\theta=\frac{\theta_0+r\left(\theta_0^2+\textstyle{\frac{1}{n-2}}\omega_0^2\right)}{1+2r\theta_0+r^2\left(\theta_0^2+\textstyle{\frac{1}{n-2}}\omega_0^2\right)} , \qquad \omega=\frac{\omega_0}{1+2r\theta_0+r^2\left(\theta_0^2+\textstyle{\frac{1}{n-2}}\omega_0^2\right)} .
\ee

This special case was already discussed in \cite{OrtPraPra07}. {Similarly as in the non-twisting case, we necessarily have here $\theta\neq 0$ \cite{OrtPraPra07} (one can also easily see that $\det L\neq 0$ and $\bl$ is thus non-degenerate), unless we consider Kundt spacetimes. If one has, in addition, also $\omega=0$, one is led to the class of Robinson-Trautman spacetimes \cite{PodOrt06}. In fact, the case $\omega\neq 0$ is possible only if $n$ is even \cite{OrtPraPra07}, so that in odd-dimensions non-shearing spacetimes thus reduce to either the Kundt or the Robinson-Trautman class.}

\subsection{Case of a matrix $L$ satisfying the ``optical constraint''} 

\label{subsec_OC}

A special case of a matrix $L$ admitting aligned principal shear and twist directions arises when $L$  satisfies the so called ``optical constraint'' \cite{OrtPraPra09}, i.e. (dropping the matrix indices) 
\be
 [S,A]=0 , \qquad A^2=S^2-{\cal F}S ,
 \label{GolSac_inv}
\ee
where ${\cal F}$ can be fixed by taking the trace (note, in particular, that $L$ is thus a normal matrix). Considering matrices $L$ satisfying this special property is motivated by the fact that this includes several important classes of vacuum solutions, namely spacetimes of type III and N \cite{Pravdaetal04}, Kerr-Schild metrics \cite{OrtPraPra09} and general asymptotically flat spacetimes with a multiple WAND \cite{OrtPraPra09b}.  In particular, in four dimensions the optical constraint is equivalent to the shearfree property (except when $A=0$ and, simultaneously, rank$(S)=1$).

One can easily see that under the above assumptions the matrix $L_{ij}$ takes the block diagonal form (cf.~also \cite{OrtPraPra09})
\beqn
 L_{ij}=\left(\begin {array}{cccc} \fbox{${\cal L}_{(1)}$} & & &  \\
  & \ddots & & \\ 
 & & \fbox{${\cal L}_{(p)}$} & \label{L_general} \\
 & & & \fbox{$\begin {array}{ccc} & & \\ \ \ & \tilde{\cal L} \ \ & \\ & & \end {array}$}
  \end {array}
 \right) . 
 \eeqn
The first $p$ blocks are $2\times 2$ and the last block $\tilde{\cal L}$ is a $(n-2-2p)\times(n-2-2p)$-dimensional diagonal matrix. They are given by (after appropriately rescaling $r$, see \cite{OrtPraPra09} for details)
\beqn
 & & {\cal L}_{(\mu)}=\left(\begin {array}{cc} s_{(2\mu)} & A_{2\mu,2\mu+1} \nonumber \\
 -A_{2\mu,2\mu+1} & s_{(2\mu)} 
\end {array}
 \right) \qquad (\mu=1,\ldots, p) , \\
  & & s_{(2\mu)}=\frac{r}{r^2+(a^0_{(2\mu)})^2} , \qquad A_{2\mu,2\mu+1}=\frac{a^0_{(2\mu)}}{r^2+(a^0_{(2\mu)})^2} , \label{s_A} \\
  & &  \tilde{\cal L}=\frac{1}{r}\mbox{diag}(\underbrace{1,\ldots,1}_{(m-2p)},\underbrace{0,\ldots,0}_{(n-2-m)}) , 
 \label{diagonal}
\eeqn
with $0\le 2p\le m\le n-2$. The integer $m$ denotes the rank of $L_{ij}$, so that $L_{ij}$ is non-degenerate when $m=n-2$.

For certain purposes it turns out to be very convenient to define the complex combination
\be
 \rho_{(\mu)}\equiv s_{(2\mu)}+iA_{2\mu,2\mu+1}=\frac{1}{r-ia^0_{(2\mu)}} .
\ee
This satisfies the compact Sachs equation 
\be
 D\rho_{(\mu)}=-\rho_{(\mu)}^2 ,\label{Sachs}
\ee
which generalizes the standard Sachs equation of the four-dimensional theory \cite{Sachs61,NP,penrosebook2,Stephanibook} (except that our $\rho_{(\mu)}$ differs by a sign from the standard NP symbol).

In the rest of the paper we will study spacetimes for which there is {at most} one non-zero block (for $\mu=1$, say). In such a situation we can drop block-indices and introduce the more compact notation $s\equiv s_{(2)}=L_{22}=L_{33}$, $A\equiv A_{23}=L_{23}=-L_{32}$, $\rho\equiv\rho_{(1)}=s+iA$. If $s=0$ then also $A=0$ \cite{OrtPraPra07}, i.e., one has the trivial solution $L=0$ corresponding to Kundt spacetimes. On the other hand, for non-zero expansion $s\neq 0$ one finds
\be
 \rho\equiv s+iA=\frac{1}{r-ia_0} ,
 \label{rho2block}
\ee
which determines the expansion scalar $\theta=2s/(n-2)$ and the twist $\omega=\sqrt{2}A$ of $\bl$.

Let us emphasize that all the above results hold for a geodetic WAND, not necessarily a multiple one, under the only assumption $R_{00}=0$ on the energy-momentum content of the spacetime. In the following we will however consider more special situations and in particular restrict to {geodetic} multiple WANDs of Einstein spacetimes.

\section{Type N spacetimes}

\label{sec_N}

In the previous section we have determined the $r$-dependence of the optical matrix $L_{ij}$ for a wide class of spacetimes. This matrix plays a special role since it enters the Bianchi identities that need to be integrated in order to find the $r$-dependence of the Weyl tensor. The full set of Bianchi identities has been presented in \cite{Pravdaetal04} (cf. also \cite{Durkeeetal10}). In the following we will use those containing $D$-derivatives, i.e., derivatives in the direction of the geodetic multiple WAND $\bl$. We start our analysis by considering type N Einstein spacetimes. The case of type III spacetimes is similar but technically more involved, and will be dealt with in the next section.

According to \cite{Pravdaetal04}, one has to consider the two possible cases $L_{ij}=0$ and $L_{ij}\neq 0$ (the results 
 of  \cite{Pravdaetal04}, relying on the Bianchi identities, were obtained in the case $\Lambda=0$, but hold also for Einstein spacetimes since, for these, $R_{abcd;e}=C_{abcd;e}$ -- the same applies in the next section in the type III case).

\subsection{Kundt spacetimes}

The condition $L_{ij}=0$ defines Kundt spacetimes (i.e., $\bl$ is non-expanding, shearfree and twistfree). In this case the Bianchi eq.~(23) of \cite{Pravdaetal04} (derived from (B.4,\cite{Pravdaetal04})) reduces to
\be
 D\Psi_{ij}=0 ,
\ee
so that
\be
 \Psi_{ij}=\Psi^0_{ij} ,
\ee
does not depend on $r$.  The amplitude of the gravitational field is unchanged as one moves along light rays (the multiple WAND), which is a ``plane wave''-like behavior. By performing a space rotation, one can thus always align one's frame to the ``polarization axes'', i.e., to the eigendirections of $\Psi^0_{ij}$.

We note in addition that Einstein spacetimes of type N belonging to the Kundt class are VSI (if $\Lambda=0$) or CSI (if $\Lambda\neq 0$) spacetimes (i.e., spacetimes for which all scalar invariants constructed from the Riemann tensor and its covariant derivatives are either vanishing or constant) \cite{Coleyetal04vsi,ColHerPel06}\footnote{The ``VSI part'' of this statement has been proven in \cite{Coleyetal04vsi} in the case $\Lambda=0$. When $\Lambda\neq0$ (i.e., for the ``CSI part'') the proof goes essentially unchanged, since the Bianchi identities 
and most of the Ricci identities which one needs are unaffected by  the cosmological term. Even though for $\Lambda\neq 0$ $N_{ij}$ is not a balanced scalar,
Lemma 4 in \cite{Coleyetal04vsi} still holds and the Weyl  (but not the Riemann) tensor and its derivatives are also balanced and thus all its invariants of all orders as well as mixed invariants with the Ricci tensor vanish.
The only difference from the vacuum case  is that one can now construct non-zero constant invariants using contractions of the Ricci tensor $R_{ab}=g_{ab}R/n$ (see also a remark at the end of Section~1 of \cite{Pravdaetal02} in the four-dimensional case). A similar comment will apply also later in the case of type III Einstein spacetimes of the Kundt class and will not be repeated there.\label{foot_CSI}} and no physically useful information can thus be extracted from their invariants. Finally, it may be also worth recalling the result of \cite{OrtPraPra09} that, in vacuum (with $\Lambda=0$), Kundt solutions of type N coincide with non-expanding Kerr-Schild spacetimes.

\subsection{Expanding spacetimes}

When $L_{ij}$ does not vanish identically the results of \cite{Pravdaetal04} implies that $L$ is normal and in particular of the special form obeying the ``optical constraint'' discussed in section~\ref{subsec_OC}. In an adapted frame (which we take to be parallelly transported, cf.~section~\ref{subsubsec_parall}) it has the only non-zero components $s\equiv s_{(2)}=s_{(3)}=(n-2)\theta/2$, and $A_{23}=-A_{32}=A$, as given in~(\ref{rho2block}).

Further, it follows from \cite{Pravdaetal04} that in such a frame $\Psi_{ij}$ has the only non-zero components\footnote{We observe that in the canonical form given in \cite{Pravdaetal04} one has $\Psi_{23}=0$. However, the frame used in that paper was not, in general, parallelly transported. We will comment more on this point shortly.}
\be  
 \Psi_{33}=-\Psi_{22}, \qquad \Psi_{23}=\Psi_{32}
 \label{Phij_N} .
\ee

Bianchi equation (23) in \cite{Pravdaetal04} can thus be compactly written as
\BE
 D(\Psi_{22}+i\Psi_{23})=-\rho(\Psi_{22}+i\Psi_{23}) ,
\EE
with the simple solution
\be
 \Psi_{22}+i\Psi_{23}=(\Psi_{22}^0+i\Psi_{23}^0)\rho .
 \label{Psi_ij_N}
\ee

Under a rotation in the $\bm^{(2)}$-$\bm^{(3)}$ plane such that the frame is still parallelly transported along $\bl$ we have $\Psi_{22}+i\Psi_{23}\to e^{i\alpha_0}(\Psi_{22}+i\Psi_{23})$, hence the $\Psi_{23}$ (or $\Psi_{22}$) component can be set to zero if and only if $a_0=0$, i.e. if the twist vanishes. In that case one has simply $\Psi_{22}=\Psi_{22}^0/r$ (or $\Psi_{23}=\Psi_{23}^0/r$) and 
such frame is thus aligned with the eigenframe of $\Psi_{ij}$  (cf.~also \cite{PraPra08}) or rotated by $45^o$.
In the twisting case this is not possible and the eigendirections of $\Psi_{ij}$ will spin with respect to the parallelly propagated frame, i.e., the effect of twist is to ``mix'' the two polarizations as one proceeds along the rays of the gravitational field (indeed, in four dimensions $\Psi_{22}$ and $\Psi_{23}$ correspond to the real and imaginary part of $\Psi_4$, respectively, i.e. the well-known ``$+$'' and ``$\times$'' polarization modes).

{Eq.~(\ref{Psi_ij_N}) can be employed, for instance, to characterize the asymptotic behavior of the Weyl tensor (whose leading term clearly behaves as $1/r$, as in four dimensions). In addition, knowing the $r$-dependence of the Weyl tensor can also be useful for studying possible spacetime singularities. Namely, one can} analyze the simplest non-trivial
curvature invariant {admitted by} type N spacetimes in four \cite{BicPra98} and higher \cite{Coleyetal04vsi} dimensions, i.e.,
\BE
 I_{N} \equiv C^{a_1 b_1 a_2 b_2  ; c_1 c_2} C_{a_1 d_1 a_2  d_2 ; c_1 c_2} C^{e_1 d_1 e_2 d_2 ;f_1 f_2} C_{e_1 b_1 e_2 b_2  ; f_1 f_2} .
 \label{InvN}
\EE
It can be shown\footnote{This was done in \cite{Coleyetal04vsi} for vacuum spacetimes with $\Lambda=0$, {but it can  be easily extended to the case $\Lambda\neq0$.}} that for type N Einstein spacetimes $I_N$ is proportional {(via a numerical constant)} to
\be
I_N\propto\left[ (\Psi_{22})^2 + (\Psi_{23})^2  \right]^2  (s^2+A^2)^4 = \frac{\left[ (\Psi_{22}^0)^2 + (\Psi_{23}^0)^2  \right]^2 }{(r^2+a_0^2)^6} .
\ee
If $a_0$ {(which generically is a function of coordinates other than $r$)} vanishes at some spacetime points, then there will be a curvature singularity at $r=0=a_0$. In particular, this always occur in the non-twisting case \cite{PraPra08}. Further {(``$r$-independent'')} singularities may arise from a possible singular behaviour of the function $(\Psi_{22}^0)^2 + (\Psi_{23}^0)^2$ (see appendix~\ref{AppN} for specific examples). {Note also that $I_N\to0$ for $r\to\infty$, i.e. far away along the multiple WAND.}

\section{Type III spacetimes}

\label{sec_III}

Similarly as for type N, also in type III vacuum spacetimes the matrix $L_{ij}$ has, in an adapted parallelly propagated frame, the only non-zero components $s\equiv s_{(2)}=s_{(3)}=(n-2)\theta/2$ and $A_{23}=-A_{32}=A$ \cite{Pravdaetal04},\footnote{More precisely, this has been proven in full generality for all $n>4$ non-twisting solutions of type III and for all $n=5$ solutions of type III (in Appendix~C of \cite{Pravdaetal04}). For twisting type III solutions with $n>5$ it was assumed that a ``generality'' condition on the components $\Psi_{ijk}$ holds and that $\Psi_i\neq 0$ (Section~4 of \cite{Pravdaetal04}). We do not consider possible ``exceptional'' cases in this paper. (Note also that the possibility rank$(S)=1$ was not discussed explicitly in section~4 of \cite{Pravdaetal04}. However, it can be shown from equations~(69), (71), (74)--(77) and (80) therein  that such case indeed cannot occur.)} given in~(\ref{rho2block}) {(the fact that an adapted frame can be taken to be parallelly transported follows from section~\ref{subsubsec_parall}, as in the type N case)}.

In order to fully determine the $r$-dependence of the Weyl tensor we shall integrate the Bianchi equations (B.1), (B.4), (B.6) and (B.9) of \cite{Pravdaetal04}, which contain $D$-derivatives of negative boost weight Weyl components. These equations (given later in an appropriate context) contain, in addition to $L_{ij}$, also the Ricci rotation coefficients $L_{1i}$, $L_{i1}$ and $\M{i}{j}{k}$, and the derivative operator $\delta_i$. We thus first need to fix the $r$-dependence of these quantities. Together with $L_{11}$ these will also determine the behaviour of the metric coefficients.

\subsection{Ricci identities}

The assumption that $\bl$ is the multiple WAND of type III spacetimes and that the frame is parallelly transported greatly simplifies the general form of the Ricci identities given in \cite{OrtPraPra07}. Recalling that the $r$-dependence of $L$ is given in (\ref{rho2block}), we can use the results given in Appendix~D of \cite{OrtPraPra09}.\footnote{Up to a small difference: Appendix~D of \cite{OrtPraPra09} studied Ricci-flat spacetimes of type II (or more special), while in this section we consider Einstein spacetimes of type III.} It will also be convenient to divide the space indices $i,j,k$ into two groups, {corresponding to the non-vanishing and vanishing block of $L_{ij}$, respectively,} i.e. 
\be 
 p, q, o=2, 3, \qquad v, w,z=4,5,\ldots,n-1 . 
\label{indices} 
\ee
 
The relevant coefficients thus read
\beqn
 & & L_{12}+iL_{13}=(l_{12}+il_{13})\rho , \qquad L_{1w}=l_{1w} , \label{L1i} \\
 & & L_{21}+iL_{31}=(l_{21}+il_{31})\bar\rho , \qquad L_{w 1}=l_{w 1} \label{Li1} , \\
 & & \M{i}{j}{2}+i\M{i}{j}{3}=(\m{i}{j}{2}+i\m{i}{j}{3})\rho , \qquad \M{i}{j}{w}=\m{i}{j}{w} , \label{Mijk} 
\\
 & & L_{11}=\mbox{Re}[(l_{12}+il_{13})(l_{21}-il_{31})\rho]+\left[\frac{R}{n(n-1)}-l_{1w}l_{w1}\right]r+l_{11} , \label{L11}
\eeqn
where, hereafter, for brevity lowercase symbols $l_{1i}$, $l_{i1}$, $\m{i}{j}{k}$ and $l_{11}$ denote quantities independent of $r$.
Because of the index symmetries of $\M{i}{j}{k}$ \cite{Pravdaetal04}, we require $\m{i}{j}{k}+\m{j}{i}{k}=0$ for any $i,j,k=2,\ldots,n-1$.

\subsection{Commutators and $r$-dependence of the metric}

In order to  determine the radial dependence of the derivative operators $\delta_i$ and $\Delta$
let us take the affine parameter $r$ as one of the coordinates and $x^A$  any set of ($n-1$) scalar functions (which need not be further specified for our purposes) such that $(r,x^A)$ is a well-behaved coordinate system. 
Then the directional derivatives (when acting on scalars) take the form 
\be
 D=\pa_r , \qquad \Delta=U\pa_r+X^A\pa_A , \qquad \delta_i=\omega_i\pa_r+\xi^A_i\pa_A ,
 \label{derivatives}
\ee
where $\pa_A\equiv\pa/\pa x^A$.

The $r$-dependence of these can be determined using the following commutators \cite{Coleyetal04vsi}
\beqn
 & & \Delta D-D\Delta=L_{11}D+L_{i1}\delta_i , \label{comm_DelD} \\
 & & \delta_i D-D\delta_i=L_{1i}D+L_{ji}\delta_j . \label{comm_dD}
\eeqn

Applying (\ref{comm_dD}) on $r$ we get $D\omega_i=-L_{1i}-L_{ji}\omega_j$,
which, using (\ref{rho2block}) and (\ref{L1i}), leads to
\be
 \omega_2+i\omega_3=(\omega^0_2+i\omega^0_3)\rho-(l_{12}+il_{13})r\rho , \qquad \omega_w=-l_{1w}r+\omega_w^0.
 \label{omegai}
\ee

Similarly, acting with (\ref{comm_dD}) on $x^A$ gives $D\xi_i^A=-L_{ji}\xi^A_j$, so that
\be
 \xi^A_2+i\xi^A_3=(\xi^{A0}_2+i\xi^{A0}_3)\rho , \qquad \xi^A_w=\xi^{A0}_w .
 \label{xiA}
\ee

Applying (\ref{comm_DelD}) on $x^A$ leads to $DX^A=-L_{i1}\xi^A_i$, from which (using (\ref{xiA})) 
\be
 X^A=\mbox{Re}[(l_{21}-il_{31})(\xi^0_2+i\xi^0_3)\rho]-l_{w1}\xi_w^{A0}r+X^{A0} .
 \label{XA}
\ee

Applying (\ref{comm_DelD}) on $r$ gives $DU=-L_{11}-L_{i1}\omega_i$. Using also (\ref{Li1}), (\ref{L11}) and (\ref{omegai}) we obtain
\be
 U=\mbox{Re}\big[[\omega^0_2+i\omega^0_3-r(l_{12}-il_{13})](l_{21}-il_{31})\rho\big]-\left[\frac{R}{2n(n-1)}-l_{1w}l_{w1}\right]r^2-(l_{11}+l_{w1}\omega^0_w)r+U^0 .
 \label{U}
\ee

Note that the above expressions for the Ricci coefficients and the derivative operators have been derived under the assumption $\rho\neq 0$ (i.e., $L_{ij}\neq0$). When this does not hold, there is no need to distinguish between two types of indices as in (\ref{indices}) and the corresponding expressions can be obtained from the above results simply by dropping all quantities containing indices 2 or 3.

Let us observe at this point that, since in the above coordinates 
\BEA
\ell^a=\delta^a_r , \qquad n^a&=U\delta^a_r+X^A\delta^a_A , \qquad m^a_{(i)}=\omega_i \delta^a_r +\xi^A_{i}\delta^a_A, 
\EEA
the $r$-dependence of all components of the frame vectors is now known. This automatically also fixes the radial dependence of the metric~(\ref{metricIII}). For the contravariant components we explicitly have  
\be
 g^{rr}=2U+\omega_i\omega_i , \qquad g^{rA}=X^A+\omega_i\xi^A_i , \qquad g^{AB}=\xi^A_i\xi^B_i ,
\ee
together with~(\ref{U}), (\ref{omegai}), (\ref{XA}) and (\ref{xiA}). 
The covariant components can be found by imposing the orthonormality relations among the frame vectors, which gives
\beqn
 & & g_{rr}=0, \qquad X^Ag_{rA}=1, \qquad \xi^A_i g_{rA}=0, \nonumber \\
 & & 2U+X^AX^Bg_{AB}=0 , \qquad \omega_i+X^A\xi^B_ig_{AB}=0,  \qquad \xi^A_i\xi^B_jg_{AB}=0 .
\eeqn
The explicit form of the covariant coefficients can be worked out more conveniently after further information about the line-element is specified and, possibly, other adapted coordinates are defined, and we will not discuss this any further here (see, e.g., \cite{PraPra08} for the non-twisting case).

We now proceed with determining the $r$-dependence of the Weyl tensor. As in the type N case, let us discuss the two possible cases $L_{ij}=0$ and $L_{ij}\neq 0$ separately.

\subsection{Kundt spacetimes}

When $L_{ij}=0$ the Bianchi equations (B.6) (or, equivalently, (B.9)) and (B.4) of \cite{Pravdaetal04} take the form
\BEA
D\Psi_{ijk}&=&0,\label{B6K}\\
2D\Psi_{ij}&=&\delta_j\Psi_{i}+\Psi_{i}(L_{1j}-L_{j1})+2\Psi_{jki}L_{k1}+\Psi_{k}\Mk_{ij}. \label{B4K}
\EEA

Direct integration of the first of these gives
\be
 \Psi_{ijk}=\Psi_{ijk}^0, \qquad \Psi_{i}=\Psi_{i}^0=2\Psi_{ijj}^0 .
 \label{KundtIIIijk}
\ee

We now discuss eq.~(\ref{B4K}). Since $\rho=0$ for Kundt spacetimes there is not need to introduce two types of indices
 and for the Ricci coefficients we have simply
\be
 L_{1i}=l_{1i} , \qquad L_{i1}=l_{i 1} , \qquad \M{i}{j}{k}=\m{i}{j}{k} , 
 \label{Ricci_Kundt}
\ee
while the derivative operator $\delta_i$ reads 
\be
 \delta_i=(-l_{1i}r+\omega_i^0)\pa_r+\xi^{A0}_i\pa_A .
 \label{delta_Kundt}
\ee

Using (\ref{KundtIIIijk}), (\ref{Ricci_Kundt}), and (\ref{delta_Kundt}) the integration of~(\ref{B4K}) gives 
\be
 \Psi_{ij}=\frac{1}{2}\left[\xi_j^{A0}\Psi^0_{i,A}+\Psi^0_i(l_{1j}-l_{j1})+2\Psi_{jki}^0l_{k1}+\Psi^0_k\m{k}{i}{j}\right]r+\Psi^0_{ij} .
 \label{KundtIIIij}
\ee

Furthermore, since $\Psi_{ii}=0=\Psi_{[ij]}$ the integration constants appearing above must satisfy
\beqn
& & \Psi^0_{ii}=0 , \qquad \xi_i^{A0}\Psi^0_{i,A}+\Psi^0_i(l_{1i}-2l_{i1}+\m{i}{j}{j})=0,  \\
  & & \Psi^0_{[ij]}=0 , \qquad 
\xi_{[j}^{A0}\Psi^0_{i],A}+\Psi^0_{[i}(l_{|1|j]}-l_{j]1})+\Psi_{jik}^0l_{k1}+\Psi^0_k\m{k}{[i}{j]}=0 .
\eeqn

{The above equations~(\ref{KundtIIIijk}) and  (\ref{KundtIIIij}) thus fully describe the $r$-dependence of the Weyl tensor for type III Einstein spacetimes of the Kundt class, in agreement with \cite{PraPra08} (where  an adapted frame such that $l_{1i}-l_{i1}=0$ was used).} Note that, in contrast to the type N case, now Weyl components of boost weight $-2$ do in general depend on $r$. This is the typical peeling-off of Kundt spacetimes (here restricted to type III) and is well-known also in four dimensions \cite{Sachs61}. 
{As discussed for the type~N, also Einstein metrics of type III that belong to the Kundt family fall in the VSI or CSI class \cite{Coleyetal04vsi,ColHerPel06}.}

\subsection{Expanding spacetimes}

Bianchi equations (B.1), (B.9), (B.6) and (B.4) of \cite{Pravdaetal04} read
\BEA
D \Psi_{i}&=&-2\Psi_{k}L_{ki},\label{B1}\\
D\Psi_{jki}&=&\Psi_{i}A_{jk}+\Psi_{kli}L_{lj}-\Psi_{jli}L_{lk},\label{B9}\\
2D\Psi_{ijk}&=&-\Psi_{i}L_{jk}+\Psi_{j}L_{ik}-2\Psi_{ijl}L_{lk},\label{B6}\\
2D\Psi_{ij}&=&-2\Psi_{ik}L_{kj}+\delta_j\Psi_{i}+\Psi_{i}(L_{1j}-L_{j1})+2\Psi_{jki}L_{k1}+\Psi_{k}\Mk_{ij}.\label{B4}
\EEA

We now study the above differential equations for various index combinations (recall~(\ref{indices})).

\subsubsection{Components of boost weight $-1$}

In terms of the two index sets~(\ref{indices}), eq.~(\ref{B1}) can be conveniently  rewritten as $D(\Psi_2+i\Psi_3)=-2\rho(\Psi_2+i\Psi_3)$ and $D\Psi_w=0$, so that
\be
 \Psi_2+i\Psi_3=(\Psi^0_2+i\Psi^0_3)\rho^2 , \qquad \Psi_w=\Psi^0_w . \label{Psii}
\ee

Let us  consider (\ref{B9}) and (\ref{B6}) for the components  $\Psi_{wvp}$. These reduce, respectively, to $D\Psi_{wvp}=0$ and $D(\Psi_{wv2}+i\Psi_{wv3})=-\rho(\Psi_{wv2}+i\Psi_{wv3})$, whose only common solution is clearly (since $\rho\neq 0$ here) 
\be
 \Psi_{wvp}=0 .
\ee

Before integrating the remaining Bianchi equations containing the operator $D$, we take advantage of the fact that for type III spacetimes some other Bianchi identities become purely algebraic. These are eqs.~(B.7), (B.11) and (B.16) of \cite{Pravdaetal04}, where a detailed analysis can be found. In particular, from (B.7) and (B.16) one can derive
eq.~(58,\cite{Pravdaetal04}), which reads
\be
\theta (n-2)\Psi_{ijk}+4S_{[i|s}\Psi_{sk|j]}-2S_{sk}\Psi_{ijs}+2S_{[i|k}\Psi_{|j]}=0 . \label{B58}
\ee
Since $\theta\not= 0$, using this with  $\{i,j,k\}=\{v,w,z\}$, $\{2,3,w\}$, $\{p,w,v\}$, $\{w,p,q\}$ we get, respectively (recall also $\Psi_{\{ijk\} }=0 $)
\BEA
 \Psi_{vwz}=0 , \qquad \Psi_{23w}=\Psi_{w32}-\Psi_{w23}=0 , \qquad \Psi_{pwv}=0, \qquad 
\Psi_{w}=0  . \label{rcePsiwpq}
\EEA
In addition, from $\Psi_i=2\Psi_{ijj}$ the last two equations give
\be
 \Psi_{233}=\frac{1}{2}\Psi_2, \quad  \Psi_{322}=\frac{1}{2}\Psi_3 ,\qquad \Psi_{w33}=-\Psi_{w22} .
\ee

Since the $r$-dependence of $\Psi_{23p}$ is thus now determined by $\Psi_p$, the only remaining equation to be solved is
eq.~(\ref{B6}) for $\Psi_{w22}$ ($=-\Psi_{w33}$) and $\Psi_{w23}$. This can be written as
\be
 D(\Psi_{w22}+i\Psi_{w23})=-\rho(\Psi_{w22}+i\Psi_{w23}) ,
\ee 
with solution
\be
 \Psi_{w22}+i\Psi_{w23}=\rho\left(\Psi_{w22}^0+i\Psi_{w23}^0\right) .  \label{Psiwpq}
\ee

This fixes the $r$-dependence of all boost weight $-1$ Weyl components  for type III Einstein spacetimes. One can check that all boost weight $+1$ and $0$ Bianchi equations 
given in \cite{Durkeeetal10} (thus, in particular, eqs.~(\ref{B1})--(\ref{B6}) and {the above mentioned algebraic equations}) are now satisfied.

\subsubsection{Constraints on the Ricci rotation coefficients}

Before we proceed with fixing  the $r$-dependence of the boost weight $-2$ Weyl components, it turns out that suitable Ricci identities will lead to considerable simplifications useful in the following calculations. Let us consider Ricci equation (11k,\cite{OrtPraPra07}), which for a geodetic null congruence $\bl$ in type III Einstein spacetimes reduces to
\be
 \delta_{[j|} L_{i|k]} =  L_{1[j|} L_{i|k]}+ L_{i1} L_{[jk]}+L_{il}\M{l}{[j}{k]}
+ L_{l[j|}\M{l}{i|}{k]} . \label{11k}
\ee

Considering the various equations obtained for $i,j=p,q$ and $k=w$ and using~(\ref{rho2block}), (\ref{L1i})--(\ref{Mijk}), (\ref{derivatives}), (\ref{omegai}) and (\ref{xiA}), one finds
\be
 \m{w}{2}{2}=\m{w}{3}{3}=\omega^0_w , \qquad \m{w}{2}{3}=-\m{w}{3}{2}=-l_{1w}a_0-\xi^{A0}_wa_{0,A} .
 \label{mwpq}
\ee

Next, for $i,j,k=o,p,q$ one gets
\be
 i(\xi^{A0}_2+i\xi^{A0}_3) a_{0,A}=-(\omega_2^0+i\omega_3^0)+ia_0\left[-(l_{12}+il_{13})+2(l_{21}+il_{31})\right ] ,
 \label{11k_opq}
\ee
and for $i=w$, $j=q$ and $k=z$ 
\be
 \m{2}{w}{z}=0=\m{3}{w}{z} .
 \label{mpwz}
\ee
Finally, for $i=w$, $j=2$ and $k=3$ we find $\m{w}{2}{3}=l_{w1}a_0$ so that, by~(\ref{mwpq}),
\be
 \m{w}{2}{3}=-\m{w}{3}{2}=l_{w1}a_0 , \qquad \xi^{A0}_wa_{0,A}=-a_0(l_{w1}+l_{1w}) . 
 \label{mw23}
\ee

Other index combinations do not contain any further information. We also note that all the remaining Ricci identities \cite{OrtPraPra07} contain also Ricci rotation coefficients that do not appear explicitly in the Weyl tensor components and thus we do not consider those here.

\subsubsection{Components of boost weight $-2$}

As the last step, the $r$-dependence of boost-weight $-2$ components of the Weyl tensor can now be determined from eq.~(\ref{B4}).
Using the above results for $\Psi_i$ and $\Psi_{ijk}$, it is convenient to study various cases with different values of the indices.

For $i,j=w,z$, recalling (\ref{Mijk}) and (\ref{mpwz}), eq.~(\ref{B4}) becomes simply
\be
 D\Psi_{wz}=0 , 
\ee
so that
\be
\Psi_{wz}=\Psi^0_{wz}, \qquad \Psi^0_{[wz]}=0 , \label{Psi_wz}
\ee
where the latter condition follows from $\Psi_{wz}=\Psi_{zw}$.

Next, for $i=w$ and $j=2,3$ eq.~(\ref{B4}) can be written as
\be
2D(\Psi_{w2}+i\Psi_{w3})=-2(\Psi_{w2}+i\Psi_{w3})\rho+\Psi_2(\M{2}{w}{2}+i\M{2}{w}{3})+\Psi_3(\M{3}{w}{2}+i\M{3}{w}{3}) .
\ee
Using (\ref{Psii}), (\ref{Mijk}), (\ref{mwpq}) and (\ref{mw23}), this leads to
\be
 \Psi_{w2}+i\Psi_{w3}=
 \rho(\Psi_{w2}^0+i\Psi_{w3}^0) +{\cal P}^0_w \rho^2
 , \qquad {\cal P}^0_w=\frac{1 }{2}(\omega^0_w+ia_0l_{w1})(\Psi_{2}^0+i\Psi_{3}^0), \label{Psi_w23}
\ee
with $\Psi_{[w2]}^0+i\Psi_{[w3]}^0=0$.

We also observe that for $i=w$, $j=2,3$ the antisymmetric part of (\ref{B4}) gives (since $\Psi_{[ij]}=0$)
\BEA
  & & \delta_w(\Psi_2+i\Psi_3)= -\rho^2\left[(l_{1w}-l_{w1}+i\m{2}{3}{w})(\Psi_{2}^0+i\Psi_{3}^0)
 +2(\Psi_{w2}^0+i\Psi_{w3}^0)+2(\Psi_{w22}^0+i\Psi_{w23}^0)(l_{21}-il_{31})\right] \nonumber \\
 & & \qquad\qquad\qquad\qquad\qquad\qquad {}-2\rho^3(\omega^0_w+ia_0l_{w1})(\Psi_{2}^0+i\Psi_{3}^0) , 
 \label{const23}
\EEA
from which (with~(\ref{derivatives}), (\ref{omegai}), (\ref{xiA}) and (\ref{mw23}))
\beqn
 & & {}-\xi^{A0}_w(\Psi_{2}^0+i\Psi_{3}^0)_{,A}=(\Psi_{2}^0+i\Psi_{3}^0)(3l_{1w}-l_{w1}+i\m{2}{3}{w})+2(l_{21}-il_{31})(\Psi^0_{w22}+i\Psi^0_{w23}) \nonumber \\
 & & \qquad\qquad\qquad\qquad\qquad\qquad {}+2(\Psi^0_{w2}+i\Psi^0_{w3}) . \label{const1} 
\eeqn

Finally, we have to consider (\ref{B4}) in the case $i,j=2,3$. The corresponding equations can be compactly rearranged as
\beqn
 & & 2D(\Psi_{22}+\Psi_{33})-(\delta_2-i\delta_3)(\Psi_2+i\Psi_3)=  (\Psi_2+i\Psi_3)\big[(L_{12}-iL_{13})-2(L_{21}-iL_{31})\nonumber \\ 
 & & \hspace{6cm} {}+(\M{2}{3}{3}+i\M{2}{3}{2})\big]-2\bar\rho(\Psi_{22}+\Psi_{33}),  \label{w-2first} \\
 & & 2D(\Psi_{22}-\Psi_{33}+2i\Psi_{23})-(\delta_2+i\delta_3)(\Psi_2+i\Psi_3)= (\Psi_2+i\Psi_3)\big[(L_{12}+iL_{13})+(-\M{2}{3}{3}+i\M{2}{3}{2})\big] \nonumber \\
 & & \hspace{6cm}  {}-4L_{w1}(\Psi_{w22}+i\Psi_{w23})-2\rho(\Psi_{22}-\Psi_{33}+2i\Psi_{23}). 
 \label{w-2second}
\eeqn

Using~(\ref{derivatives}), (\ref{omegai}), (\ref{xiA}) and (\ref{Psii}) we find the needed transverse derivatives, i.e., 
\beqn
 & & (\delta_2+i\delta_3)(\Psi_2+i\Psi_3)=2\rho^4(\Psi_2^0+i\Psi_3^0)[i(\xi^{A0}_2+i\xi^{A0}_3) a_{0,A}-(\omega_2^0+i\omega_3^0)+r(l_{12}+il_{13})] \nonumber \label{delta1} \\ 
 & & \hspace{6cm} {}+\rho^3(\xi^{A0}_2+i\xi^{A0}_3)(\Psi_2^0+i\Psi_3^0)_{,A} , \\
 & & (\delta_2-i\delta_3)(\Psi_2+i\Psi_3)=2\rho^3\bar\rho(\Psi_2^0+i\Psi_3^0)[i(\xi^{A0}_2-i\xi^{A0}_3) a_{0,A}-(\omega_2^0-i\omega_3^0)+r(l_{12}-il_{13})] \nonumber \\ 
 & & \hspace{6cm} {}+\rho^2\bar\rho(\xi^{A0}_2-i\xi^{A0}_3)(\Psi_2^0+i\Psi_3^0)_{,A}. \label{delta2}
\eeqn

After substituting (\ref{delta1}) into (\ref{w-2second}), by direct integration we find
\beqn
 \Psi_{22}-\Psi_{33}+2i\Psi_{23}=
  \rho^3{\cal A}^0+\rho^2{\cal B}^0+\rho{\cal C}^0+\rho r{\cal D}^0 ,\label{Psi_22m33}
\eeqn
where (using also (\ref{11k_opq})) 
\BEA
{\cal A}^0&=& (\Psi_2^0+i\Psi_3^0)\left[(\omega_2^0+i\omega_3^0)-ia_0(l_{21}+il_{31})\right] ,\nonumber\\
{\cal B}^0&=&-\frac{1}{2}\left[(\Psi_2^0+i\Psi_3^0)[3(l_{12}+il_{13})+i(\m{2}{3}{2}+i\m{2}{3}{3})]
 +(\xi^{A0}_2+i\xi^{A0}_3)(\Psi_2^0+i\Psi_3^0)_{,A}\right] , \nonumber \\ 
{\cal C}^0&=&\Psi_{22}^0-\Psi_{33}^0+2i\Psi_{23}^0 , \label{A_B_C_D} \\ 
{\cal D}^0&=&-2l_{w1}(\Psi_{w22}^0+i\Psi_{w23}^0).  \nonumber
\EEA
{Note that here ${\cal C}^0$ is the only new integration ``constant''.}

Similarly, by substituting (\ref{delta2}) into (\ref{w-2first}), one has 
\be
 \Psi_{22}+\Psi_{33}=\rho\bar\rho {\cal F}^0+\bar\rho{\cal G}^0,\hspace{2cm} \label{Psi_{22p33}}
 \label{P22+P33} 
\ee
where
\BEA
{\cal F}^0&=&-\frac{1}{2}\left[(\Psi_2^0+i\Psi_3^0)[3(l_{12}-il_{13})+(\m{2}{3}{3}+i\m{2}{3}{2})-2(l_{21}-il_{31})]
 +(\xi^{A0}_2-i\xi^{A0}_3)(\Psi_2^0+i\Psi_3^0)_{,A}\right] , \nonumber \\ 
{\cal G}^0&=&\Psi_{22}^0+\Psi_{33}^0 ,
\label{E_F_G}
\EEA
and ${\cal G}^0$ is the new integration constant.

However, since $\Psi_{ii}=0$, we have $\Psi_{22}+\Psi_{33}=-\Psi_{ww}=-\Psi^0_{ww}$.
This must now be compatible with (\ref{P22+P33}), from which we get 
\be
 \Psi^0_{ww}=0 , \qquad \Psi_{22}^0+\Psi_{33}^0=0 , \qquad {\cal F}^0=0  ,
 \label{const22}
\ee
so that
\be
 \Psi_{22}+\Psi_{33}=0=\Psi_{ww} .
\ee

\subsubsection{Summary and discussion}

The above results can now be conveniently summarized as follows. First, a number of Weyl components vanish identically, namely
\beqn
 & & \Psi_w=0, \qquad \Psi_{ijw}=0,  \qquad \Psi_{wz2}=0=\Psi_{wz3} , \\
 & & \Psi_{22}+\Psi_{33}=0 . \label{Psi_22+}
\eeqn

The $r$-dependence of the non-zero components is given by
\beqn
 & & \Psi_2+i\Psi_3=2(\Psi_{233}+i\Psi_{322})=\rho^2(\Psi^0_2+i\Psi^0_3) , \label{Psi_i} \\
 & & \Psi_{w22}+i\Psi_{w23}=-\Psi_{w33}+i\Psi_{w32}=\rho(\Psi_{w22}^0+i\Psi_{w23}^0) , \label{Psi_wpq} \\ 
 & & \Psi_{wz}=\Psi^0_{wz} \qquad\qquad  (\Psi^0_{[wz]}=0=\Psi^0_{ww}) , \label{Psi_wz2} \\
 & & \Psi_{w2}+i\Psi_{w3}=\rho(\Psi_{w2}^0+i\Psi_{w3}^0) + \rho^2{\cal P}_w^0 , \qquad\qquad (\Psi_{[w2]}^0=0=\Psi_{[w3]}^0) \label{Psi_wp} \\
 & & 2(\Psi_{22}+i\Psi_{23})=\rho^3{\cal A}^0+\rho^2{\cal B}^0+\rho{\cal C}^0+\rho r{\cal D}^0 , \label{Psi_22-} 
\eeqn
where the various integration ``constants'' satisfy (\ref{Psi_w23}), (\ref{A_B_C_D}), and (\ref{const22}) with (\ref{E_F_G}), along with the constraints~(\ref{11k_opq}), (\ref{mw23}) and  (\ref{const1}).

As $r\to\infty$, one can observe a different fall-off behavior for different components and therefore a peeling-like behavior. There exist components of boost weight $-1$ both with $1/r^2$ (eq.~\ref{Psi_i}) and $1/r$ (eq.~\ref{Psi_wpq}) leading terms. The slower fall-off described by the latter equation can be qualitatively understood as due to the fact that there is no expansion along the ``$w$-directions''. As for boost weight $-2$, in general there are components that are asymptotically constant in $r$ (eqs.~(\ref{Psi_wz2}) and (\ref{Psi_22-})) and components that fall off as $1/r$ (eq.~(\ref{Psi_wp})). Again, the asymptotically ``constant'' terms are due to the non-expanding extra-dimensions.

{There are several special subcases that may be worth mentioning. First, for the special subtype III$(a)$, which is invariantly defined by the condition $\Psi_i=0$ \cite{Coleyetal04}, one obtains the simplifying conditions ${\cal P}_w^0={\cal A}^0={\cal B}^0=0$ and, by~(\ref{const1}), $\Psi^0_{w2}+i\Psi^0_{w3}=-(l_{21}-il_{31})(\Psi^0_{w22}+i\Psi^0_{w23})$ (further simplification can be achieved by using a residual frame freedom, see below).}   Next, also in the non-twisting case ($\rho=1/r$) the above expressions~(\ref{Psi_i})--(\ref{Psi_22-}) become much simpler and were given already in \cite{PraPra08} (in particular, thanks to~(\ref{11k_opq}) one gets ${\cal A}^0=0$, so that the $\rho^3$ term of (\ref{Psi_22-}) disappears).
Finally, one can compare the above results with the well-known asymptotic behaviour in four dimensions (cf., e.g., \cite{Talbot69,Lind74,NewTod80}). In that case, in our notation, indices $v,w,z$ do not exist (since $i,j,k=2,3$ only) and eqs.~(\ref{Psi_i}) and (\ref{Psi_22-}) encode all the information, corresponding, respectively, to the complex Newman-Penrose scalars $\Psi^{(NP)}_3\sim 1/r^2$ and $\Psi^{(NP)}_4\sim 1/r$ (note that ${\cal D}^0=0$ in four dimensions). There are no non-expanding extra-dimensions and thus no terms with a slower fall-off.

{In the general case}, for certain applications the asymptotic behavior of the Weyl components~(\ref{Psi_i})--(\ref{Psi_22-}) can in fact be visualized more clearly by taking a series expansion. Using $\rho=\sum_{m=1}^\infty(ia_0)^{m-1}r^{-m}$, $\rho^2=\sum_{m=1}^\infty m(ia_0)^{m-1}r^{-(m+1)}$ and $\rho^3=\frac{1}{2}\sum_{m=1}^\infty m(m+1)(ia_0)^{m-1}r^{-(m+2)}$, up to the leading and sub-leading terms one finds
\beqn
 & & \Psi_{2}+i\Psi_{3}=\left( \frac{\Psi_{2}^0}{r^2}-\frac{2a_0\Psi_{3}^0}{r^3}\right)
+i\left( \frac{\Psi_{3}^0}{r^2}+\frac{2a_0\Psi_{2}^0}{r^3}\right) +{\cal O}(r^{-4}), \\
 & & \Psi_{w22}+i\Psi_{w23}=\left( \frac{\Psi_{w22}^0}{r}-\frac{a_0\Psi_{w23}^0}{r^2}\right)
+i\left( \frac{\Psi_{w23}^0}{r}+\frac{a_0\Psi_{w22}^0}{r^2}\right) +{\cal O}(r^{-3}), \\
 & & \Psi_{w2}+i\Psi_{w3}=\left(\frac{\Psi_{w2}^0}{r}+\frac{\omega^0_w\Psi^0_2-a_0(2\Psi_{w3}^0+l_{w1}\Psi_{3}^0)}{2r^2}\right) +i\left(\frac{\Psi_{w3}^0}{r}+\frac{\omega^0_w\Psi^0_3+a_0(2\Psi_{w2}^0+l_{w1}\Psi_{2}^0)}{2r^2}\right)\nonumber \\
 & & \hspace{10cm} {}+{\cal O}(r^{-3}), \\
 & & \Psi_{22}+i\Psi_{23}=\left(-l_{w1}\Psi_{w22}^0+\frac{\Psi_{22}^0+a_0l_{w1}\Psi_{w23}^0}{r}\right)+i\left(-l_{w1}\Psi_{w23}^0+\frac{\Psi_{23}^0-a_0l_{w1}\Psi_{w22}^0}{r}\right)+{\cal O}(r^{-2}) , \nonumber \\ 
 & & 
\eeqn
{while still $\Psi_{wz}=\Psi_{wz}^0$. This clearly demonstrates, in particular, how the presence of twist mixes up various polarization modes. Simpler expressions can be obtained by performing specific frame transformations, as briefly discussed below.}

{Similarly as for the type N, the $r$-dependence of the Weyl tensor can also be used to discuss the possible presence of curvature singularities.} The simplest non-trivial curvature invariant for expanding type III Einstein spacetimes is 
\cite{Pravda1999,Coleyetal04vsi} 
\BE
I_{{III}} = C^{a_1 b_1 a_2 b_2;e_1} C_{a_1 c_1 a_2 c_2;e_1} C^{d_1 c_1 d_2 c_2;e_2} C_{d_1 b_1 d_2 b_2;e_2}.\label{invIII}
\EE
It can be shown that \cite{Coleyetal04vsi}
\BEA
 & & \hspace{-.35cm} I_{{III}} \propto (s^2+A^2)^2[9(\Psi_i\Psi_i)^2+27(\Psi_i\Psi_i)({\Psi_{w22}}{\Psi_{w22}} + \Psi_{w23}\Psi_{w23})
+28({\Psi_{w22}}{\Psi_{w22}} + \Psi_{w23}\Psi_{w23})^2]\nonumber\\
 & & \hspace{-.35cm} = \left[ 9 \frac{\left[(\Psi_{2}^0)^2+(\Psi_{3}^0)^2\right]^{2}}{(r^2+a_0^2)^2} +27\frac{ \left[(\Psi_{2}^0)^2+(\Psi_{3}^0)^2\right]
({\Psi_{w22}^0}{\Psi_{w22}^0} + \Psi_{w23}^0\Psi_{w23}^0) }{r^2+a_0^2}+28(\Psi_{w22}^0\Psi_{w22}^0 + \Psi_{w23}^0\Psi_{w23}^0)^2 \right] \nonumber \\ 
 & & \qquad\qquad\qquad {}\times \frac{1}{(r^2+a_0^2)^4} .
\label{Inv3}
\EEA
As in the type N case, there may be curvature singularities localized at points where $r^2+a_0^2=0$, which may or may not exist, in general (but they always do in the non-twisting case \cite{PraPra08}). Additional  singularities may also arise from a possible singular behaviour of $\Psi_{2}^0 $, $\Psi_{3}^0$, $\Psi_{w22}^0$ and  
$\Psi_{w23}^0$ (see appendix~\ref{AppIII} for specific examples).

\subsubsection{Frame freedom}

The results above have been obtained using a generic parallelly transported frame and therefore hold in any such frame. For certain purposes it may be desirable to simplify some expressions by using the freedom to perform null rotations
\BE
 \bl\to\bl, \qquad \bn\to\bn+z_i\bm^{(i)} -\pul z_kz_k\bl , \qquad \bm^{i}\to\bm^{(i)} -z_i\bl ,
 \label{nullrot}
\EE
where $Dz_i=0$ in order for the new frame to be still parallelly transported.

For instance, once can set to zero the Ricci rotation coefficients $L_{12}+iL_{13}$ or $L_{21}+iL_{31}$ (by taking $z_2+iz_3=-(l_{12}+il_{13})$ or $z_2+iz_3=-(l_{21}+il_{31})$, respectively), or one may want to set to zero certain Weyl components of boost weight $-2$ (for type III spacetimes components of boost weight $-1$ are invariant under~(\ref{nullrot})). Namely, one can transform away the term ${\cal B}^0$ in (\ref{Psi_22-}) by taking $2(z_2+iz_3)(\Psi^0_2+i\Psi^0_3)={\cal B}^0$ (note that if $\Psi^0_2+i\Psi^0_3=0$ then ${\cal B}^0$ is automatically zero and there is no need to perform any null rotations). Alternatively, if $\Psi_{w22}^0+i\Psi_{w23}^0\neq0$ one can set  $\Psi_{w2}^0+i\Psi_{w3}^0$  in (\ref{Psi_wp}) to zero by taking $(z_2-iz_3)(\Psi_{w22}^0+i\Psi_{w23}^0)=\Psi_{w2}^0+i\Psi_{w3}^0$. Additionally, there are null rotations in the $w$-directions. If $\Psi_{w22}^0+i\Psi_{w23}^0\neq0$ one can take $4z_w(\Psi_{w22}^0+i\Psi_{w23}^0)=-{\cal C}^0$
 so as to have, in the new frame, ${\cal C}^0=0$ in (\ref{Psi_22-}) (note that this is not possible in four dimensions since $\Psi_{w22}^0+i\Psi_{w23}^0=0$ in that case). Or one can set ${\cal P}^0_w=0$ in (\ref{Psi_wp}) by taking $z_w(\Psi^0_2+i\Psi^0_3)=2{\cal P}^0_w$ (if $\Psi^0_2+i\Psi^0_3=0$ then ${\cal P}^0_w=0$ already in the original frame). 

Furthermore, spatial rotations can also be used to simplify the form of certain Weyl components. For example,
a spatial rotation in the $\bm_{(2)}$-$\bm_{(3)}$ plane adds an arbitrary ($r$-independent) phase to all the above non-zero components (except for $\Psi_{wz}$, which is unchanged) and can thus be used to set to zero the imaginary part of the corresponding integration constants (in particular, in the non-twisting case $\rho$ is real and one can thus align one's frame to the ``polarization'' of such components). Next, one can use rotations in the planes defined by the ``non-expanding'' directions $\bm_{(w)}$ to, e.g., align the frame to the direction defined by $\Psi_{w22}$ or $\Psi_{w23}$, etc., or to diagonalize $\Psi_{wz}$. The most convenient way how to use the frame freedom may depend on the specific spacetime under consideration and its possible symmetries.

\section{Concluding remarks}

After presenting some general results about the Sachs equations (section~\ref{sec_Sachs}), we studied specific features of Einstein spacetimes of type N and III in arbitrary higher dimensions. This is a natural extension of previous studies such as \cite{Pravdaetal04,PraPra08} and partly complements, in different respects, other works either by the present authors or by others, e.g.,  \cite{OrtPraPra09b,PodZof09}. In particular, by explicitly determining the $r$-dependence of the Weyl tensor we were able to discuss several physical properties of the general families of solutions of type N/III, either with or without expansion, and to compare these with their well-known four-dimensional counterparts. The results of this paper also represent a first step towards the exact integration of the full Newman-Penrose or GHP equations for such spacetimes, which will be studied elsewhere. In the following appendix the discussed results are illustrated by presenting some explicit solutions that, to our knowledge, have not been given before.

\section*{Acknowledgments}

We are grateful to Pawe\l{} Nurowski for useful email correspondence and for providing us with some references.
This work has been supported by {research plan No AV0Z10190503 and research grant GA\v CR P203/10/0749}.

\renewcommand{\thesection}{\Alph{section}}
\setcounter{section}{0}

\section{Some explicit expanding spacetimes}

\renewcommand{\theequation}{A.\arabic{equation}}
\setcounter{equation}{0}

In the main text we have studied properties of general Einstein spacetimes of type N and III in higher dimensions, for an arbitrary value of the cosmological constant. While Kundt solutions are similar to their four-dimensional counterparts and several explicit examples are already known \cite{Coleyetal03,Obukhov04,ColHerPel06,Coleyetal06,PodZof09,OrtPraPra09}, not many type N/III spacetimes with $L_{ij}\neq 0$ have been found. To our knowledge, in fact, the only such examples have been obtained (for $\Lambda=0$)\footnote{An Einstein space which is the direct product of non-Ricci-flat Einstein spaces also contains Weyl components of boost weight~0 \cite{PraPraOrt07} and thus can not be of type N/III.\label{foot_direct}} as a direct product of a four-dimensional type N/III Ricci-flat spacetime 
with an Euclidean Ricci-flat space (see, e.g., \cite{PraPra08}). The lack of less trivial examples is partly due to the fact that, contrary to the four-dimensional case, they are necessarily shearing \cite{Pravdaetal04} and therefore they do not show up, e.g., in the Robinson-Trautman family \cite{PodOrt06}.

In this appendix we present a few examples of such solutions (both non-twisting and twisting) which are {\em not} direct products. They are in fact warped products and solve the vacuum Einstein equations $R_{ab}=2\Lambda g_{ab}/(n-2)$ with a possible cosmological constant ({which can take an arbitrary value, at least in the non-twisting examples}). As a reader familiar with four-dimensional exact solutions may easily note, these spacetimes have been constructed by appropriate ``warping'' of four dimensional type N/III solutions (see, e.g.,  sections~13.3.3, 28.1, 28.4, 29.1--29.4 of \cite{Stephanibook} and section~19.2 of \cite{GriPodbook}; {some of the original papers are also quoted below in the appropriate context}). 

In fact, all the considered metrics can be written in the form (cf.~\cite{Brinkmann25})
\be
 \d s^2=\frac{1}{f(z)}\d z^2+f(z)\d\sigma^2 ,  
 \label{ansatz}
\ee
where 
\be
 f(z)=-\lambda z^2+2dz+b , \qquad \lambda=\frac{2\Lambda}{(n-1)(n-2)} ,
\ee
$b$ and $d$ are constant parameters,\footnote{In fact, $\lambda$ is the only physically meaningful free parameter contained in $f(z)$ (as one can always redefine $z\to\alpha z+\beta$) but for convenience we will generally keep also $b$ and $d$ unspecified.} and $\d\sigma^2$ is a {Lorentzian} Einstein spacetime of dimension $n-1$ with Ricci scalar 
\be
 R_\sigma=(n-1)(n-2)(\lambda b+d^2).
\label{ricci-n-1}
\ee 
This metric will be specified in the following and will characterize the properties of the full spacetime $\d s^2$. We observe that the latter is a direct product only in the special case of a constant $f(z)$, i.e. $\lambda=0=d$ (with $b>0$). In order to have a Lorentzian signature for $\d s^2$, we require $f(z)>0$, which may restrict  possible parameter values and (possibly) the range of $z$. Namely, {since $f(z)$ has real roots if and only if $R_\sigma \geq 0$},
when $R_\sigma \le 0$ we require $\lambda<0$ ($R_\sigma = 0$ admits also $\lambda=0$, but this case simply corresponds to a direct product), while for $R_\sigma>0$ any sign of $\lambda$ (including $\lambda=0$) is admitted, at least for suitable values of $z$.

\subsection{Einstein spacetimes of type N}
\label{AppN}

\subsubsection{Non-twisting case}

\label{subsubsec_N_nontwist}

 Using the above general ansatz~(\ref{ansatz}), one can obtain five-dimensional type N Einstein spacetimes with an arbitrary value of the cosmological constant $\lambda$ by taking $\d\sigma^2$ to be the general four-dimensional expanding and non-twisting type N Einstein metric with {a possibly non-zero (four-dimensional) Ricci scalar $R_\sigma=12(\lambda b+d^2)$}. This was given in \cite{GarPle81} (see also \cite{GriPodbook} and references therein, in particular for a transformation to the standard Robinson-Trautman coordinates) and reads
\be
 \d\sigma^2= -2 \psi \d u \d r +2 r^2 (\d x^2+ \d y^2)-2 r (2 r f_1+\epsilon x) \d u \d x - 2 r (2 r f_2+ \epsilon y) \d u \d y + 2 (\psi B+A)\d u^2,
\ee
with 
\BEA
A &=& \frac{1}{4}\epsilon^2(x^2+y^2)+\epsilon (f_1 x+f_2y)r+(f_1^2+f_2^2)r^2, \\
B &=& -\frac{1}{2} \epsilon-r \partial_x f_1+\frac{1}{24} R_\sigma r^2 \left(1+\frac{1}{2} \epsilon (x^2+y^2) \right), \\
\psi &=& 1+\frac{1}{2} \epsilon (x^2+y^2),
\EEA
where $\epsilon=\pm 1$ or $0$ and the functions $f_1=f_1(x,y)$  and $f_2=f_2(x,y)$ are subject to
\BE
\partial_x f_1 = \partial_y f_2, \quad \partial_y f_1 = - \partial_x f_2 .
\label{holomorphic}
\EE

In the case $\lambda b+d^2=0$ the metric $\d\sigma^2$ is Ricci-flat and the spacetime $\d s^2$ can be lifted to any higher dimensions by simply replacing $\d\sigma^2\to\d\sigma^2+\sum_\alpha(\d \tilde z_\alpha)^2$. One can obtain a higher dimensional solution also in the case $\lambda b+d^2\neq0$, however in a bit more complicated way (a simple direct product will not work as it will introduce Weyl components of boost weight zero, cf.~footnote~\ref{foot_direct}). For simplicity, in the following analysis we will restrict to the $n=5$ case.

The geodetic multiple WAND is given by 
\be
 \bl=\pa_r ,
\ee
with $r$ being an affine parameter along the corresponding null geodesics. We can then choose a parallelly transported frame in the form 
\beqn
 & & \bn=- \frac{ 1}{\psi f(z)} \partial_u - \left[\left(
\frac{\lambda b + d^2}{f(z)}-\frac{\lambda}{2}\right)r^2 - \frac{  \epsilon/2 +r \partial_x f_1}{ \psi  f(z)}  \right]  \partial_r
- \frac{   1}{    \psi f(z) } \left(f_1+ \frac{\epsilon x}{2r} \right) \partial_x \nonumber \\
& & \qquad\qquad\qquad {}-\frac{   1}{    \psi f(z) } \left(f_2+ \frac{\epsilon y}{2r} \right) \partial_y-(\lambda z-d )r \partial_z , \\
 & &  \bm_{(2)} = \frac{1}{r} \frac{1}{\sqrt{2f(z)}} \partial_x, \qquad
 \bm_{(3)} = \frac{1}{r} \frac{1}{\sqrt{2f(z)}} \partial_y,  \qquad
 \bm_{(4)} = \frac{\lambda z - d}{\sqrt{f(z)}}r\partial_r+\sqrt{f(z)}\partial_z .
\eeqn

Then the only non-vanishing Ricci rotation coefficients relevant to the discussion in the main text are (note that $L_{11}\neq 0$ but we do not need it here)
\be
 L_{22}=L_{33}=\frac{1}{r} ,\qquad 
 L_{21}+iL_{31}= \frac{- \epsilon (x+iy)}{  r \sqrt{2 f(z)} \psi}, \qquad  L_{14}=-L_{41}=-\frac{\lambda z-d}{\sqrt{f(z)}} .
\ee 

The non-vanishing {independent} components of the Weyl tensor are (after using~(\ref{holomorphic}))
\be
\Psi_{22}+i\Psi_{23}=-\frac{(\partial^3_y+i\partial^3_x)f_2}{4 f(z)^2 \psi} \frac{1}{r} .
\ee

The curvature invariant $I_N$ given in (\ref{InvN}) is therefore proportional to 
\be
 I_N\propto\frac{ \left[ \left(\partial^3_x f_2\right)^2 + \left(\partial^3_y  f_2\right)^2 \right]^2}
 { \psi^4 f(z)^8 r^{12}} . 
\ee
Similarly as in the four-dimensional case \cite{BicPra98}, in five dimensions $I_N$ diverges whenever {any of the following conditions hold:} i) $r=0$; ii) $\psi = 0$ (i.e., for $\epsilon=-1$ and $x^2+y^2=2$); iii) the quantity 
$\left(\partial^3_x f_2\right)^2 + \left(\partial^3_y  f_2\right)^2$
diverges. In five dimensions an additional  curvature singularity is also located at the roots of  $f(z)=0$, which are present iff $R_{\sigma} \geq 0$.

\subsubsection{Twisting case}

Here we present a five-dimensional twisting Einstein spacetime of type N with a negative cosmological constant $\lambda$.
This is constructed by taking $\d\sigma^2$ in~(\ref{ansatz}) to be the four dimensional type N twisting solution of Leroy \cite{Leroy1970} (but in different coordinates, cf.~also \cite{Stephanibook,Siklos81,Nurowski08}) with a negative {Ricci scalar} $R_\sigma=12(\lambda b+d^2)\equiv-4s^2$, i.e.,
\be
\d \sigma^2 =
\frac{1}{s^2 y^2} \left[ \frac{3}{2} (r^2+1) (\d x^2+\d y^2)+ \frac{1}{3} (\d x+y^3 \d u) \left[6 y \d r +y^3 (1-r^2) \d u+(13-r^2) \d x+12 r \d y \right] \right].
\label{Leroy}
\ee

The coordinate $r$ is an affine parameter along the geodetic multiple WAND $\bl = \pa_r$. 
We choose a parallelly propagated frame 
\BEA
 \bn &=& -\frac{w_1 (3 r^2-1)}{4 r y^3}  \pa_u + w_2 \pa_r -\frac{w_1}{r} \pa_x
 +w_1 \pa_y - (\lambda z-d) r \pa_z, \\
 \bm_{(2)} &=& \frac{s\sqrt{2/3}}{y^2  (r^2+1)} \frac{1}{\sqrt{f(z)}} \left(r   \pa_u + 4 y^2 r  \pa_r - y^3 r  \pa_x -y^3  \pa_y \right), \\
 \bm_{(3)} &=& \frac{s\sqrt{2/3}}{y^2  (r^2+1)} \frac{1}{\sqrt{f(z)}} \left(   \pa_u + 4 y^2   \pa_r - y^3  \pa_x + y^3 r \pa_y \right), \\
 \bm_{(4)} &=&  \frac{\lambda z-d}{\sqrt{f(z)}}r\pa_r + \sqrt{f(z)} \pa_z, 
\EEA
where 
\be
 w_1=-\frac{4s^2 r y}{3f(z) (r^2+1)}, \qquad w_2=\frac{\lambda r^2}{2} +\frac{s^2(2 r^4 +9 r^2 -25)}{6f(z)(r^2+1)}.
\ee

The non-vanishing components of the optical matrix are 
\be
 L_{22}+iL_{23}=L_{33}-iL_{32}= \frac{1}{r-i}.
\ee
The remaining relevant non-zero Ricci rotation coefficients are 
\BEA
&& L_{21}-i L_{31}=2i(\M{2}{3}{2} + i \M{2}{3}{3})=\frac{2 is}{r-i} \sqrt{\frac{2}{3f(z)}} 
 , \qquad L_{14}=-L_{41} = -\frac{\lambda z-d}{\sqrt{f(z)}}, \\
&& \M{2}{4}{2}+i\M{2}{4}{3}=-i(\M{3}{4}{2}+i\M{3}{4}{3})=\frac{-i}{r-i} \frac{\lambda z-d}{\sqrt{f(z)}} . 
\EEA
The Weyl tensor components are
\be
\Psi_{22}+i \Psi_{23} = \frac{7 i s^4}{9(r-i) f(z)^2} , 
\ee
{in agreement with the general result~(\ref{Psi_ij_N}).} Hence for the curvature invariant $I_N$ ({\ref{InvN}}) we have
\be
 I_N\propto \frac{s^{16}}{(r^2+1)^6 f(z)^8}. 
\ee
Due to above mentioned relation $R_\sigma=-4s^2<0$, the warp function $f(z)$ has no real roots and therefore in this case
 $I_N$ is everywhere regular. In addition, the components of the Weyl tensor in the above frame (parallelly propagated along $\bl$) are also regular. See~\cite{Siklos81} for a discussion of the regularity of the four-dimensional Leroy  metric~(\ref{Leroy}).

To conclude, we note that Leroy's solution was rediscovered in \cite{Nurowski08} as a special case of a more general class of four-dimensional twisting Einstein spacetimes of type N (determined up to solving a system of two third-order ODEs). Similarly as above, these metrics can be used to construct other type N solutions in five dimensions (in particular, also Ricci-flat ones if one starts from a four dimensional geometry with a positive Ricci scalar). Finally, let us also observe that further five (or higher)-dimensional type N solutions with a negative cosmological constant can easily be constructed by taking $\d\sigma^2$ to be the well-known four dimensional type N Hauser solution \cite{Hauser74} (cf.~also \cite{Stephanibook}).

\subsection{Einstein spacetimes of type III}
\label{AppIII}

\subsubsection{Non-twisting case}

An explicit solution in $n=5$ dimensions is given by~(\ref{ansatz}) with
\be
 \d\sigma^2=\frac{r^2}{x^3}(\d x^2+\d y^2)+2\d u\d r+\left(\frac{3}{2}x+(\lambda b+d^2)r^2\right)\d u^2 ,
 \label{III_nontwist}
\ee
{which is a four-dimensional Robinson-Trautman spacetime of type III \cite{Stephanibook}.} The cosmological constant $\lambda$ can take an arbitrary value. {Similarly as in~\ref{subsubsec_N_nontwist}, when $\lambda b+d^2=0$ the above spacetime can be lifted to any higher dimensions by simply replacing $\d\sigma^2\to\d\sigma^2+\sum_\alpha(\d \tilde z_\alpha)^2$, but again in the following we will restrict to the $n=5$ case.}

A geodetic multiple WAND is given by 
\be
 \bl=\pa_r ,
\ee
while a suitable parallelly transported frame consists of 
\beqn
 & & \bn=\frac{1}{f(z)}\pa_u-\left[\left(\frac{\lambda b+d^2}{f(z)}-\frac{\lambda}{2}\right)r^2+\frac{33x}{32f(z)}\right]\pa_r-\frac{3x^2}{4f(z)r}\pa_x-(\lambda z-d)r\pa_z , \\ 
 & & \bm_{(2)}=\frac{\sqrt{x}}{\sqrt{f(z)}}\left(\frac{3}{4}\pa_r+\frac{x}{r}\pa_x\right) , \quad \bm_{(3)}=\frac{x^{3/2}}{\sqrt{f(z)}}\frac{1}{r}\pa_y , \quad \bm_{(4)}=\frac{\lambda z-d}{\sqrt{f(z)}}r\pa_r+\sqrt{f(z)}\pa_z . \nonumber
\eeqn

The only non-zero components of the optical matrix $L_{ij}$ are given by
\be
 L_{22}=L_{33}=\frac{1}{r} .
\ee

The remaining non-zero Ricci rotation coefficients relevant to the discussion in the main text are
\be
 L_{12}=L_{21}=\M{2}{3}{3}=-\frac{3\sqrt{x}}{4\sqrt{f(z)}}\frac{1}{r} , \qquad L_{14}=-L_{41}=-\frac{\lambda z-d}{\sqrt{f(z)}} .
\ee

For the non-zero Weyl tensor components one finds
\be
  \Psi_2=2\Psi_{233}=\frac{3x^{3/2}}{4f^{3/2}(z)}\frac{1}{r^2} , \qquad \Psi_{24}=\frac{3x^{3/2}(\lambda z-d)}{8f^2(z)}\frac{1}{r} ,
\ee
which are special cases of the general expressions~(\ref{Psi_i}) and (\ref{Psi_wp}).

For the curvature invariant~(\ref{invIII}) we now have
\BE
 I_{III}\propto\frac{x^6}{f^6(z)r^{12}} .
\EE
There are thus obvious curvature singularities at $r=0$ and $x\to\infty$  and, in the case  $\lambda b+d^2 \geq 0$, additional curvature singularities {are} present at the roots of $f(z)$.

{Other four-dimensional Robinson-Trautman spacetimes of type III \cite{Stephanibook} can be also used instead of~(\ref{III_nontwist}) to obtain different five-dimensional solutions.}

\subsubsection{Twisting case}

A five-dimensional solution with a negative cosmological constant is given by
\be
 f(z)=-\lambda z^2 , \qquad (\lambda<0) ,
\ee
and {by taking $\d\sigma^2$ to be the four-dimensional Ricci-flat type III metric \cite{Stephanibook}}
\beqn
 \d\sigma^2= & & \left(\frac{\Sigma}{x^3}+6y^2xw^2\right)\d x^2+\frac{\Sigma}{x^3}\d y^2+2(\sqrt{13}+1)yx^2w^2\d x\d y \nonumber \\
 & & {}-xw\d u\left(6y\d x+(\sqrt{13}+1)x\d y\right)-4yw\d r\d x+2\d u\d r+\frac{3}{2}x\d u^2 ,
 \label{III_twist}
\eeqn
where the functions $w=w(x)$ and $\Sigma=\Sigma(r,x)$ are defined as
\be
 w=ax^{(\sqrt{13}-5)/2} , \qquad \Sigma=r^2+x^6w^2 .
\ee
The constant parameter $a$ gives rise to non-zero twist, and for $a=0$ (i.e., $w=0$ and  $\Sigma=r^2$) the solution corresponding to~(\ref{III_twist}) reduces to the previous~(\ref{III_nontwist})  (in the case $b=0=d$).

As above, the coordinate $r$ is an affine parameter along the geodetic multiple WAND 
\be
 \bl=\pa_r .
\ee
A parallelly transported frame is now given by
\beqn
 & & \bn=\frac{1}{f(z)}\left(1-\frac{3ryx^2w}{2\Sigma}\right)\pa_u-\left[-\frac{\lambda}{2}r^2+\frac{33x}{32f(z)}+\frac{3(1+\sqrt{13})x^7w^2}{8\Sigma f(z)}\right]\pa_r-\frac{3rx^2}{4f(z)\Sigma}\pa_x \nonumber \\
 & & \qquad\qquad\qquad {}-\frac{3x^5w}{4f(z)\Sigma}\pa_y-\lambda zr\pa_z , \nonumber \\ 
 & & \bm_{(2)}=\frac{\sqrt{x}}{\sqrt{f(z)}\Sigma}\left[2ryxw\pa_u+\left(\frac{3}{4}\Sigma+\frac{1}{2}(1+\sqrt{13})x^6w^2\right)\pa_r
 +xr\pa_x+x^4w\pa_y\right] , \nonumber \\
 & &  \bm_{(3)}=\frac{x^{3/2}}{\sqrt{f(z)}\Sigma}\left[x^2w\left(-2y xw\pa_u+\frac{1}{2}(1+\sqrt{13})r\pa_r-x\pa_x\right)+r\pa_y\right] , \\ 
 & & \bm_{(4)}=\frac{\lambda z}{\sqrt{f(z)}}r\pa_r+\sqrt{f(z)}\pa_z . \nonumber
\eeqn

The only non-zero components of the optical matrix $L_{ij}$ are given by
\be
 L_{22}+iL_{23}=L_{33}-iL_{32}=\frac{1}{r+ix^3w} .
\ee

Other relevant non-zero Ricci rotation coefficients are
\beqn
 & & L_{12}+iL_{13}=L_{21}-iL_{31}=-i(\M{2}{3}{2}+i\M{2}{3}{3})=-\frac{3\sqrt{x}}{4\sqrt{f(z)}}\frac{1}{r+ix^3w} , \qquad L_{14}=-L_{41}=-\frac{\lambda z}{\sqrt{f(z)}} , \nonumber \\
 & & \M{2}{4}{2}+i\M{2}{4}{3}=-i(\M{3}{4}{2}+i\M{3}{4}{3})=\frac{\lambda zx^3w}{\sqrt{f(z)}}\frac{i}{r+ix^3w} .
\eeqn

The non-zero Weyl tensor components are given by
\beqn
  & & \Psi_{2}+i\Psi_{3}=2(\Psi_{233}+i\Psi_{322})=\frac{3x^{3/2}}{4f^{3/2}(z)}\frac{1}{(r+ix^3w)^2} , \\
  & & \Psi_{22}+i\Psi_{23}=\frac{3(1+\sqrt{13})x^5w}{16f^2(z)}\frac{i}{(r+ix^3w)^3} , \qquad \Psi_{24}+i\Psi_{34}=\frac{3x^{3/2}\lambda z}{8f^2(z)}\frac{r}{(r+ix^3w)^2} .
\eeqn
These can be compared with the general expressions~(\ref{Psi_i}), (\ref{Psi_22-}) and (\ref{Psi_wp}). 

For the curvature invariant (\ref{invIII}) we thus have 
\BE
 I_{III}\propto\frac{x^6}{\lambda^6z^{12}(r^2+x^6w^2)^{6}} .
\EE
Curvature singularities are thus located at $r=0$ for $x\to0$ (more in general, the regularity at $r=0=x$ depends on how this location is approached) and at $z=0$ (while for $x\to\infty$ we have $I_{III}\to 0$, as long as $a\neq 0$). 

Note that the above five-dimensional spacetime can be lifted to any higher dimensions by replacing $\d\sigma^2\to\d\sigma^2+\sum_\alpha(\d \tilde z_\alpha)^2$.

It is finally worth observing that a four-dimensional twisting spacetimes of type III with a negative cosmological constant was given in \cite{Siklos81} (see also \cite{Stephanibook}), while several Ricci-flat solutions are known, see \cite{Stephanibook} and references therein. These can all be used instead of~(\ref{III_twist}) to obtain other five-dimensional solutions.


\providecommand{\href}[2]{#2}\begingroup\raggedright\endgroup

\end{document}